\documentclass{JINST}

\title{Plans for $\overline{\sf P}${\sf ANDA} Online Computing}

\author{
Jens S\"oren Lange$^a$\thanks{Corresponding author.}, 
Dapeng Jin$^b$, 
Daniel Kirschner$^a$,
Andreas Kopp$^a$, 
Wolfgang K\"uhn$^a$, 
Johannes Lang$^a$,
Lu Li$^b$, 
Ming Liu$^a$, 
Zhen-An Liu$^b$, 
David Münchow$^a$,
Tiago Perez$^a$, 
Johannes Roskoss$^a$, 
Qiang Wang$^b$,
Hao Xu$^b$, 
and Shuo Yang$^a$\\
\llap{$^a$} II.\ Physikalisches Institut, Justus-Liebig-Universit\"at Giessen\\
Heinrich-Buff-Ring 16, 35392 Gie\ss{}en, Germany\\
\llap{$^b$} Institute of High Energy Physics, The Chinese Academy of Sciences,\\
19B Yuquan Road, Beijing 100049, China\\
E-mail: \email{soeren.lange@exp2.physik.uni-giessen.de}
\vspace*{1cm}\\
\sc\large
Presented at Workshop on fast $\breve{\rm C}$erenkov detectors,\\ 
May 11-13, 2009, Giessen, Germany\\
}

\abstract{
The $\overline{\sf P}${\sf ANDA} experiment will not use any hardware trigger,
i.e.\ all raw data are streaming in the data acquisition
with a bandwidth of $\leq$280~GB/s.
The $\overline{\sf P}${\sf ANDA} Online System is designed 
to perform data reduction 
by a factor of $\simeq$800 by reconstruction algorithms
programmed in VHDL (Very High Speed Integrated Circuit 
Hardware Description Language) on FPGAs (Field Programmable
Gate Arrays).
}


\begin{document}

\section{Introduction}

The $\overline{\sf P}${\sf ANDA} experiment at the future FAIR 
(Facility for Antiproton and Ion Resarch) facility 
at GSI Darmstadt, Germany, 
will investigate $\overline{p}$+$p$ and $\overline{p}$+$A$ 
collisions. It will be a fixed target experiment using 
a frozen hydrogen pellet target and 
a beam of $\leq$$10^{11}$ stored and cooled antiprotons
with a beam momentum $p$$\leq$15~GeV/c.
The beam momentum resolution will be $\delta$$p$/$p$$\geq$$10^{-5}$
and the luminosity $\cal{L}$$\leq$2$\times$10$^{32}$~cm$^{-2}$s$^{-1}$.
Among many other topics, the physics program will cover 
the production of charmonium states 
in the reaction $\overline{p}$$p$$\rightarrow$$\overline{c}$$c$.
If one adjusts the beam energy to resonant $J$/$\psi$ production
for one year, and assumes a duty factor of 50\%, this will 
correspond to a number of $\leq$2$\times$10$^9$ $J$/$\psi$. 
In particular, $\overline{\sf P}${\sf ANDA} will be able to measure the width 
of charmonium states in the order of $\geq$100~keV.
Other physics topics \cite{meson08} are spin physics 
(e.g.\ measurement of generalized parton distributions)
and hypernuclear physics 
(e.g.\ production of double hypernuclear nuclei). 

$\overline{\sf P}${\sf ANDA} will be one of the very few experiments worldwide 
not using any hardware trigger. All raw data will be streaming into 
the data acquisition (DAQ), and need to be filtered before being recorded 
to tape.
The reason for this approach is,  that signal events
such as charmonium events in 
$\overline{p}$$p$$\rightarrow$$\overline{c}$$c$
have a very similar event topology
compared to background events such as 
$\overline{p}$$p$$\rightarrow$$\overline{u}$$u$, $\overline{d}$$d$, $\overline{s}$$s$.
There are no straight-forward trigger criteria such as
number of charged tracks or number of neutral clusters 
in the calorimeter.
Thus, the only way of data reduction is online reconstruction 
on a farm with high computing performance.
Algorithms can be e.g.\ invariant mass reconstruction
on a particular charmonium state, 
and then applying e.g.\ a cut on a signal 
in the invariant mass in the $\overline{\sf P}${\sf ANDA} online system.

\section{The $\overline{\sf P}${\sf ANDA} Experiment}

\subsection{The $\overline{\sf P}${\sf ANDA} Detector}

One of the important tasks to be performed by the 
$\overline{\sf P}${\sf ANDA} online system
will be the online particle indentification (PID), i.e.\ assigning
a probability that a given charged track is a pion, kaon, proton,
electron or muon.
For this purpose, the data of the central 
$\overline{\sf P}${\sf ANDA} $\breve{\rm C}$erenkov detector
DRC (\underline{D}etector for internally \underline{r}eflected 
$\breve{\rm \underline{C}}$erenkov light)
plays an essential role. It is a detector of DIRC type, 
i.e.\ using internally reflected $\breve{\rm C}$erenkov light, 
consisting of 16 quartz bars (refractive index $n$=1.47) 
of thickness $d$=1.7~cm at a radius of $R$=48~cm. 
For the central tracking system, two detector options are still
under evaluation, both covering a radial range of $R$=15-41~cm:
a TPC (\underline{T}ime \underline{P}rojection \underline{C}hamber) with 135 padrows
and in total 135,169 pads of 2$\times$2 mm$^2$ size, or
a STT (\underline{S}traw \underline{T}ube \underline{T}racker)
with 4100 straw tubes with a tube radius $R$=1~cm and a tube length $L$=1.5~m,
aranged in 15 double layers. Axial or skewed arrangement
with respect to the beam axis is used, the skewed tubes 
being used for $z$ reconstruction.
As part of the charged particle tracking near the target, 
an MVD (\underline{M}icro \underline{V}ertex \underline{D}etector)
consisting of $\simeq$10$^7$ silicon pixels of size 100$\times$100~$\mu$m$^2$ and 
$\simeq$7$\times$10$^4$ strips will be implemented.
Further technical details about $\overline{\sf P}${\sf ANDA} 
are described elsewhere \cite{panda_tpr}.

\subsection{The $\overline{\sf P}${\sf ANDA} Data Acquisition System}

\label{c_daq}

With a high event rate of 
$\leq$2$\times$10$^7$ events/s and a raw event size of 4-20 kB
(average 14 kB) $\overline{\sf P}${\sf ANDA} will reach a data rate of $\leq$280~GB/s,
the same order of magnitude as LHC experiments.
As a difference, $\overline{\sf P}${\sf ANDA} will not utilize any hardware triggers, 
but all raw data will be streamed to the DAQ.
The baseline hardware platform for the $\overline{\sf P}${\sf ANDA} DAQ system 
are Compute Nodes (CN), which will be described in detail 
in Ch.~\ref{c_cn}. The CNs will run online reconstruction
algorithms programmed in VHDL on FPGAs for data reduction. 
All data digitization will be performed even in a stage
before the CNs by the frontend electronics.
Further details can be found elsewhere \cite{konorov}.

\subsection{The $\overline{\sf P}${\sf ANDA} Offline Computing System}

The $\overline{\sf P}${\sf ANDA} offline computing system is characterized
by the large amount of data to be recorded.
The final rate of events written to tape,
at a stage behind the online data reduction system,
is designed as 25~kHz.
Assuming one year of data taking with a duty factor of 50\%,
this corresponds to 3.78$\times$10$^{11}$ events.
With an estimated event size of $\simeq$4~kB for DSTs\footnote{
DSTs will be the final reduced data set to be used 
for physics analyses. They contain e.g.\ 4-momenta
of charged particles and neutral particles, 
but no reconstructed detector hit data anymore.} 
(Data Summary Tapes), this corresponds to 
$\geq$1,5 Pbyte per year, or $\simeq$378,000 DVDs.
Including not only the DSTs, but also raw data, 
Monte-Carlo simulated data, reconstructed detector hit data etc., 
an estimate for the amount of data to be stored for only the 
first year of $\overline{\sf P}${\sf ANDA} data taking will be $\simeq$11.5~Pbyte.
The offline computing will be performed
on $\simeq$2000 quad core CPUs
for reconstruction, analysis and MC production.

\subsection{The $\overline{\sf P}${\sf ANDA} Online Computing System}

\label{conline}

From the constraints of the data acquisition system on the one side
and of the offline computing system on the other side,
the requirement for the online computing system
can be defined, i.e.\ to reduce
2$\times$10$^7$ events/s raw data
to 25$\times$10$^3$ event/s to be recorded to tape.
This corresponds to a reduction factor of $\simeq$800.

\section{The HADES Experiment}

First test beams for $\overline{\sf P}${\sf ANDA} are envisaged 
for 2016. However, in particular the programming of the algorithms 
has already started. In order to be able to test online algorithmus 
already by now with real data, data from the HADES experiment were used, 
which studies dielectron events in 
$p$+$p$, $p$+$A$ and $A$+$A$ collisions, e.g.\ for investigating
the behaviour of vector mesons inside nuclear matter.
These vector mesons are detected by their decay into $e^+$$e^-$.
Therefore HADES uses a RICH 
(Ring Imaging $\breve{\rm C}$erenkov) detector for $e^+$ and $e^-$
identification. A ring finder is used online on the 
Level-2 trigger system.
Charged tracks are identified in HADES by 4 drift chambers 
of trapezoidal shape with $\simeq$30 $m^2$ of active area.
2 chambers are located in front and 2 behind a solenoid field
for momentum measurement. Each drift chamber has 6 layers
of wires, arranged in different angles for assigning a hit position
and a track direction in each chamber.
The HADES RICH detector has 55,296 readout pads of different geometrical shapes.
Signal rings induced by $e^+$ or an $e^-$ have a fixed ring radius of 4 pads.
Further details are described elsewhere \cite{hades_nim}.
Thus, several of the algorithmus for $\overline{\sf P}${\sf ANDA} 
(e.g.\ ring finder and track finder)
can be tested (with modifications) already on real data from HADES.
In addition, HADES will be upgraded in the near future, 
in order to be prepared for heavy collision systems such as $Au$+$Au$
collisions with high track multiplicity and thus higher required data
bandwidth. Therefore a new data acquisition system and Level-2 trigger system 
has been proposed based on the CN, and the algorithms could be part of
the upgraded trigger system.

\section{The Compute Node}

\label{c_cn}

\begin{figure}[hhh] \centering
\centerline{
\includegraphics[width=0.6\textwidth]{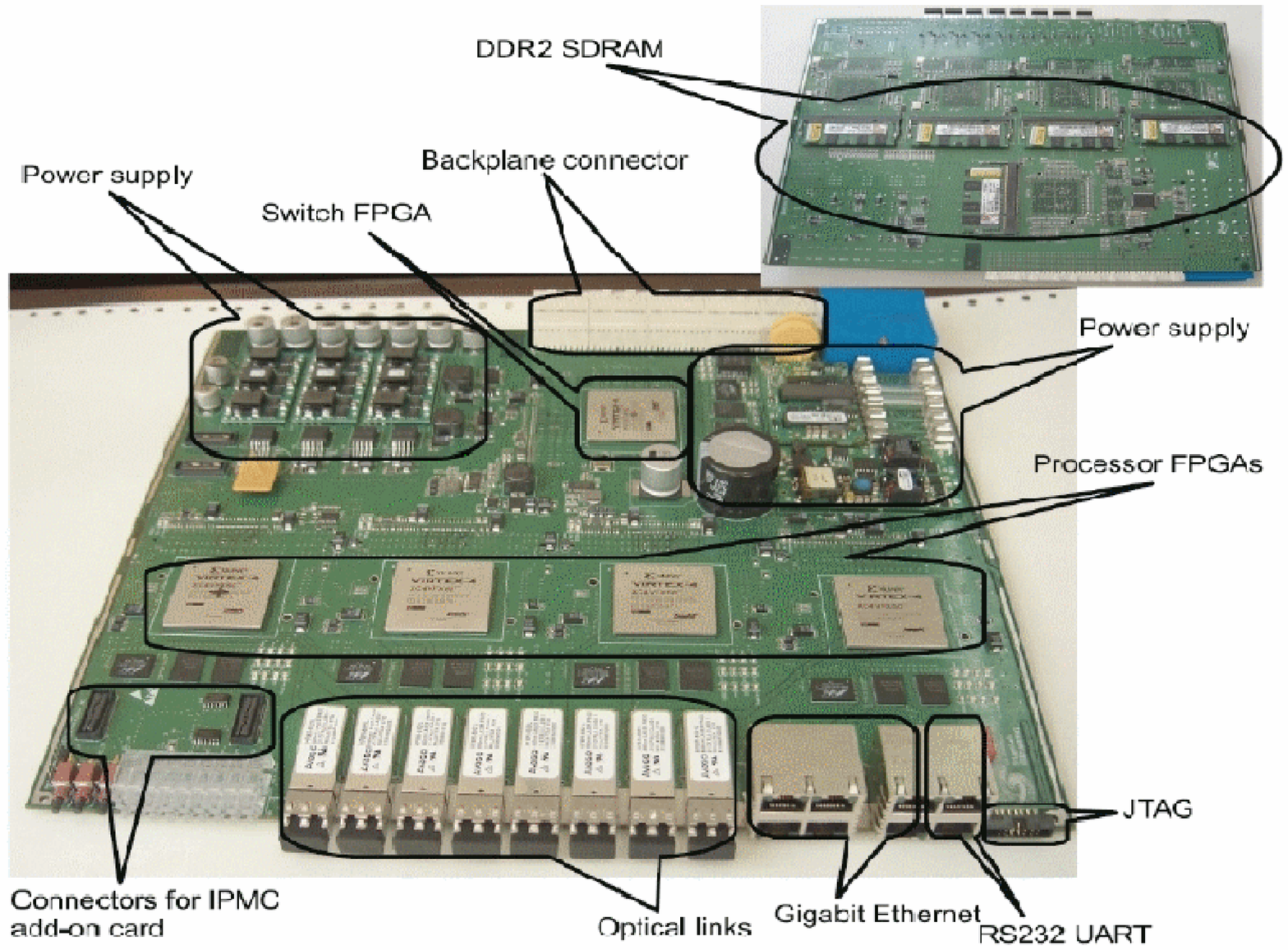}
\includegraphics[width=0.4\textwidth]{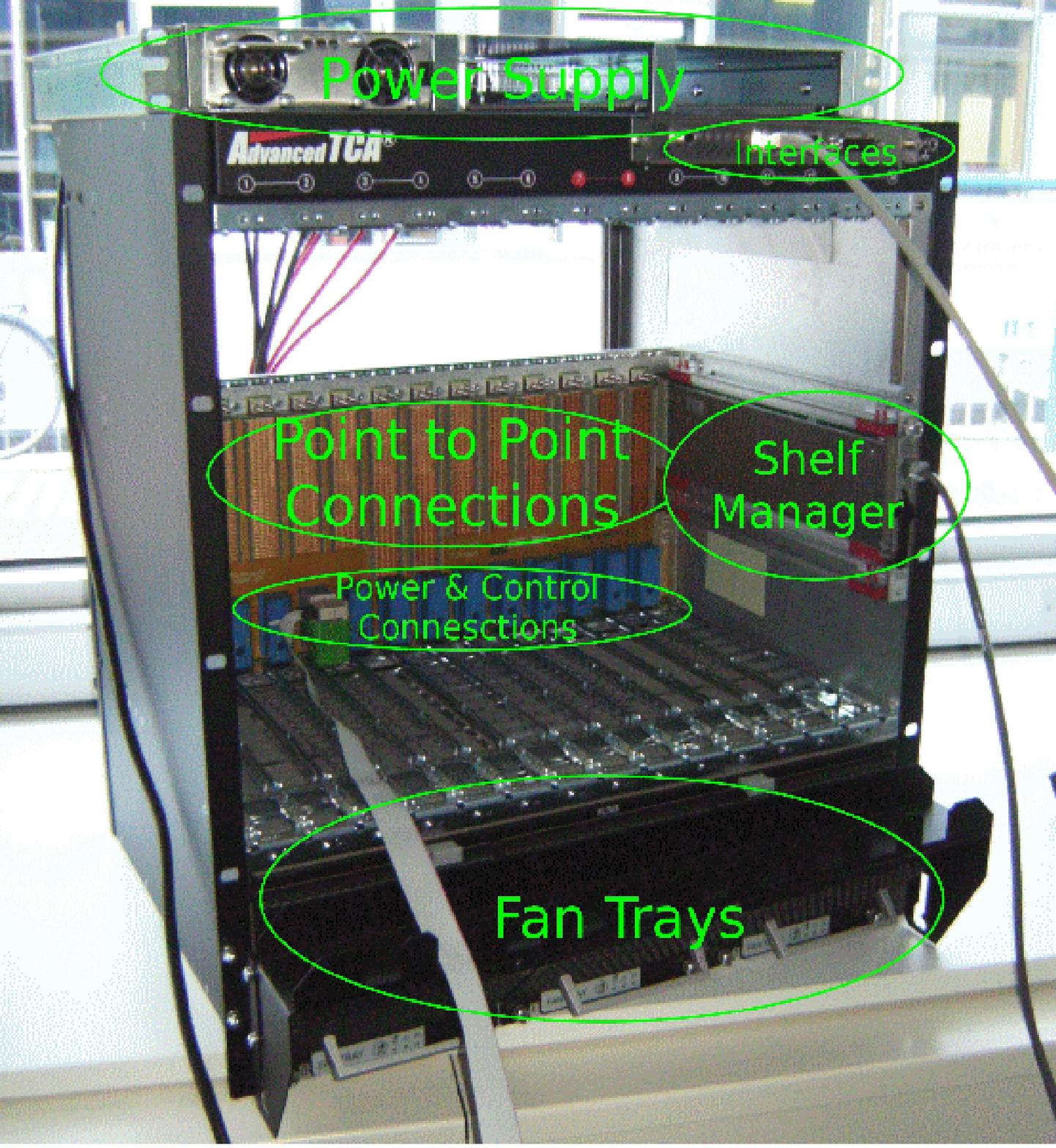} 
}
\caption{
{\it Left:} Photo of a prototype of the Compute Node (CN).
{\it Right:} Photo of an ATCA Shelf.
} 
\label{fcn}
\end{figure}

The proposed hardware unit to perform 
the online reconstruction at $\overline{\sf P}${\sf ANDA}
is the {\sc Compute Node} (CN) and is shown in Fig.~\ref{fcn}.
The 14-layer printed circuit board has been developed by IHEP Beijing
and the II.\ Physics Department of University Giessen.
Each CN has five VIRTEX-4 FX-60 FPGAs (Field Programmable Gate Arrays).
These FPGAs were chosen, as they combine 
high computing performance on the one hand
and links for high bandwidth data transfer (RocketIO)
on the other hand.
One main feature of the board design is, that all FPGAs
are connected point-to-point (see also below for details)
in order to 
{\it (a)} combine data of different regions of one detector, 
processed by different FPGAs, and 
{\it (b)} combine data of different detectors within one event
(i.e.\ event building).
The programming of the FPGAs in VHDL
is using XILINX ISE (Integrated Software Environment) Vers.\ 10.1.
As an important note for algorithm design, 
FPGAs only provide fixed\footnote{There are softcores for floating point
calculations for FPGAs available, however, the performance is not competetive
to other architectures such as GPUs (see Ch.~\ref{c_gpu}).} point arithmetics.
Thus, for any calculations such as matrix multiplications
or trigonometric functions, 
{\it (a)} the parameter range has to be fixed 
(in order to limit it into a given fixed precision range), and 
{\it (b)} lookup tables have to be used instead of calculating 
arithmetics functions.
Each Virtex-4 FX60 FPGA has two 300 MHz PowerPCs implemented as core,
however, these are only used for slow control purposes
and not for algorithms. In the current design, 
the Power PCs are booting Linux 2.6.27.
In addition, each FPGA has 2~GB of DDR2 memory attached.
The power negotiation and other slow control tasks 
between the CN and the ATCA shelf is 
based upon IPMI (Intelligent Platform Management Interface), 
implemented by an ATMEL ATmega2560 microcontroller 
on a CN add-on card \cite{thesis_lang}.
The CN is designed as a board of the ATCA 
(Advanced Telecommunications Computing Architecture)
standard.
The ATCA shelf is shown in Fig.~\ref{fcn}.
In an ATCA shelf with a full mesh backplane, 
point-to-point connections from each CN to each other CN are wired.
This avoids any bus arbitration.
In addition to the high computing performance,
the CN also provide high bandwidth interconnections.
{\it (a)} All 5 FPGAs are connected pairwise (on the board) by
one 32-bit general purpose bus (GPIO) and one full duplex RocketIO link.
{\it (b)} 4 of 5 FPGAs have two RocketIO links routed to front panel
using Multi-Gigabit Receivers (MGT) for optical links.
{\it (c)} One of the 5 FPGAs serves as a router and has 16 RocketIO links 
through the full mesh backplane 
to all the other compute nodes in the same ATCA shelf.
{\it (d)} All 5 FPGA have a Gigabit Ethernet Link routed to front panel.
With the current design, the input bandwidth in one ATCA crate
is $\leq$35~GB/s (14 CN, eight optical links each, 
operating at $\leq$2.5~Gbit/s).
The output bandwidth is $\simeq$2.6~GB/s 
(14 CN, five GB Ethernet links each, 
operating at 0.3~Gbit/s TCP performance, measured in \cite{thesis_ming}). 
All RocketIO links are currently operated with $\leq$2.5~Gbit/s,
but the upgrade to $\leq$6.5~Gbit/s is envisaged, which would
lead to even higher required reduction factors.

\section{Algorithms}

As mentioned above, 
one of the important tasks to be performed by the online reconstruction 
system at $\overline{\sf P}${\sf ANDA}
will be the online particle identification (PID).
Several subtasks have to be accomplished 
in order to achieve online PID assignment:\\
{\it (a)} An online ring finder for the DRC, 
whereas $\breve{\rm C}$erenkov photons propagate and are reflected inside 
the quartz bars, then exit the bars at the downstream end,
and generate rings in the focal plane. After applying the 
ring finder, the ring radius $R_{ring}$ is known.\\
{\it (b)} a track finder and a track fitter for charged tracks,
with hits in the MVD and STT or TPC. After the stage of the 
track fitter, the 3-momentum $\vec{p}_{track}$, and in particular the size of 
the momentum $p_{track}$=$|$$\vec{p}_{track}$$|$ of the charged track is known.\\
{\it (c)} the extrapolation of the track onto the surface of the DRC 
(in order to know, at which $z_{track}$ position the particle entered
the quartz bar).\\
{\it (d)} The $\breve{\rm C}$erenkov angle $\vartheta_{\breve{\rm C}erenkov}$ 
is a function of the two parameters $R_{ring}$ and $z_{track}$,
and will be implemented as a lookup table in the online system.\\
{\it (e)} The final PID decision will be based upon a 2-dimensional 
plot of $\vartheta_{\breve{\rm C}erenkov}$ vs.\ $p_{track}$.\\
These algorithm steps will be performed the farm of CN,
which was described in Ch.~\ref{c_cn}.
In the following, examples will be given for {\it track finder}
and {\it ring finder} algorithms. 
These algorithms are either tested with Monte-Carlo data 
for $\overline{\sf P}${\sf ANDA} or real data for HADES.

\subsection{Track Finder Algorithm for HADES}

A straight line track finder algorithm was tested with HADES data \cite{thesis_ming},
using the 2 drift chambers in front of the $B$ field, 
i.e.\ $\leq$12 fired wires out of 2110 wires define a track.
The algorithm was fully implemented on an FPGA.
The processing time of the FPGA was compared to the CPU time of C program, 
performing the same track finder task, but running on a Xeon 2.4 GHz.
For different fired wire multiplicities
$N_{wire}$=10-400 a speedup of a factor 10.8-24.3
with respect to the reference was achieved.

\subsection{Ring Finder Algorithm for HADES}

\begin{figure}[hhh] \centering
\centerline{\includegraphics[width=\textwidth]{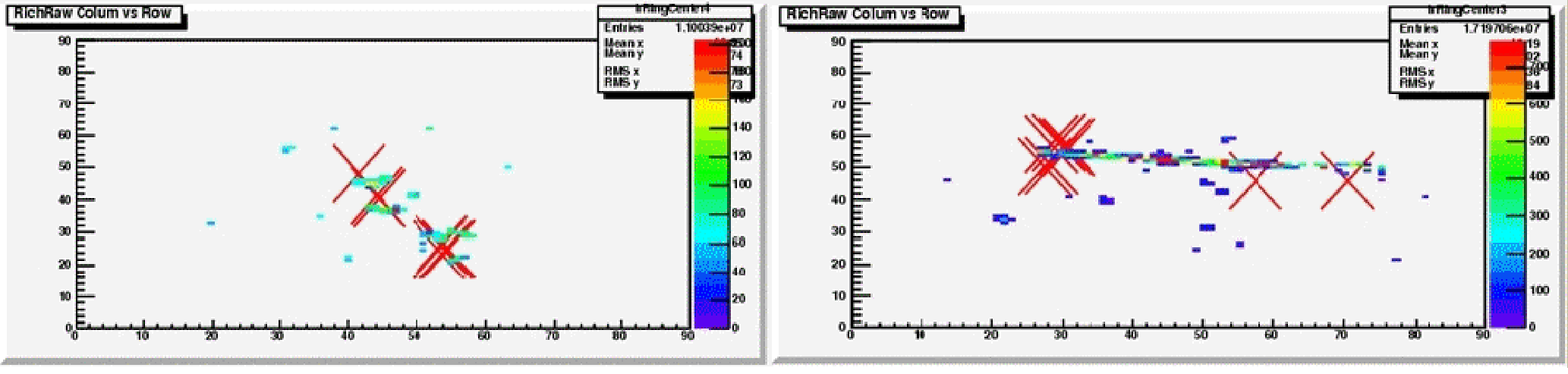}}
\caption{
Hit maps for the pad plane of the HADES RICH detector
for $^{12}$$C$+$^{12}$$C$ collisions at a beam energy of
1~AGeV. {\it Left:} Dielectron candidate event with two rings.
{\it Right:} Background event with a charged particle 
crossing the padplane. Extrapolated tracks for matching 
with ring center are shown as red crosses.
} 
\label{fring}
\end{figure}

The existing HADES online ring finder system is implemented on a VME board 
with 12 Xilinx XC4028EX FPGAs \cite{thesis_lehnert}. As such it is 
part of the HADES Level-2 trigger system \cite{thesis_traxler}
and is in operation for several years of data taking
\cite{thesis_alberica} \cite{thesis_camilla}.
For an improved algorithm, to be implemented on the CN
for the HADES upgrade project, the matching of a ring with 
a track (from the two drift chamber planes in front of  
the solenoid field) is foreseen \cite{thesis_roskoss}.
Rings are only searched in regions-of-interest in the pad plane,
given by areas of 13$\times$13 pads, centered around a pad,
which' position was found by track extrapolation.
As the RICH uses a mirror, 
reflecting the $\breve{\rm C}$erenkov light 
onto a pad plane in upstream direction,
another coordinate transformation is required by usage 
of a lookup table. The pad plane for a typical signal and 
a typical background event is shown in Fig.~\ref{fring}.
In order to quantitatively compare for the old and the new algorithm,
the enrichment factor for lepton candidates for real data is evaluated.
The enrichment factor is defined as the ratio of the 
efficiency\footnote{The efficiency is defined as the number of good positive triggers,
divided by the sum of the numbers of good positive and false negative 
triggers.}
and the reduction factor\footnote{The reduction factor 
is defined as the sum of the numbers 
of good positive triggers and false positive triggers, 
divided by the number of downscaled triggers.}
For $^{12}$C+$^{12}$C at 1~AGeV, using the new algorithm, 
the enrichment increases from 8.9 to 14.6, 
while the efficiency drops only from 93\% to 91\%.
For $^{40}$Ar+$^{40}$$[$KCl$]$ at 1.756 AGeV
with a higher track density the 
enrichment increases only from 1.7 to 2.0,
again with a minor efficiency drop from 91\% to 90\%.

\subsection{Track Finder Algorithm for $\overline{\sf P}${\sf ANDA}}

\begin{figure}[hhh] \centering
\centerline{\includegraphics[width=\textwidth]{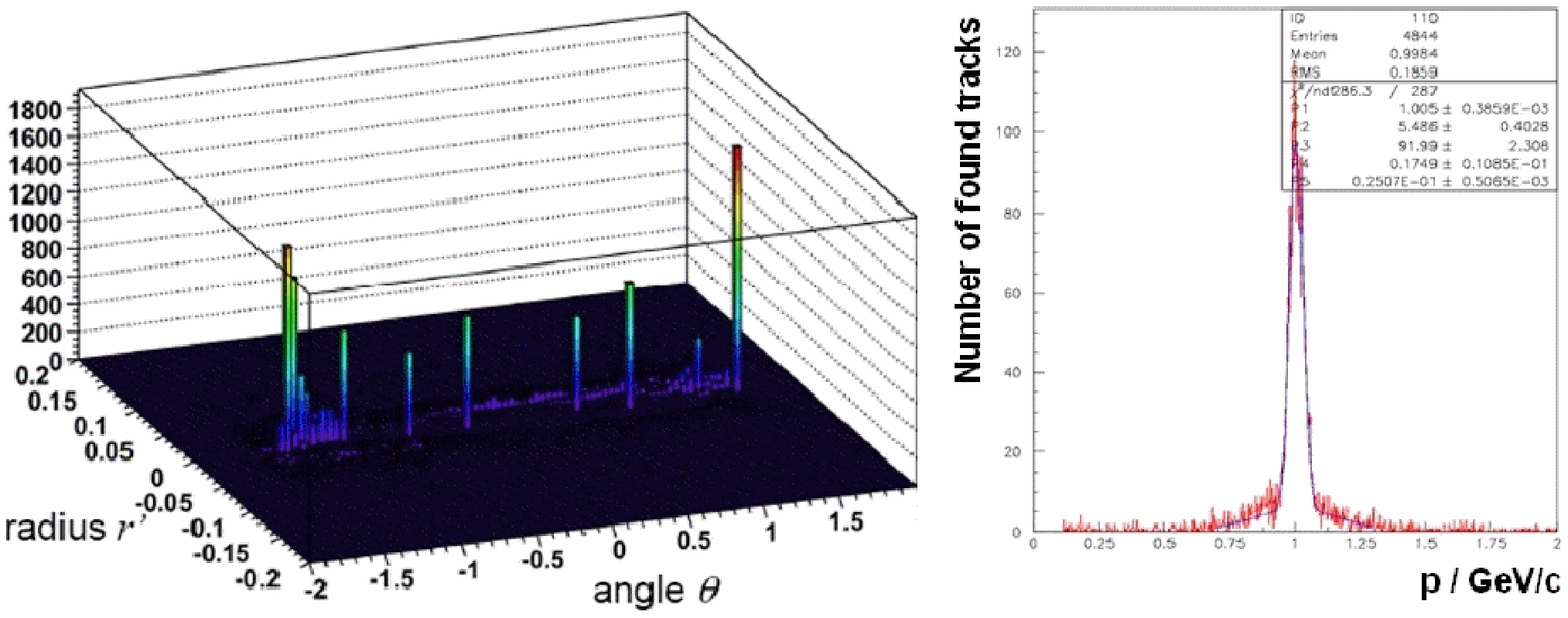}}
\caption{
{\it Left:} Hough space for 10 tracks with $p$=1~GeV/c. 
For details see text. {\it Right:} Reconstructed momentum 
for tracks with $p$=1~GeV/c. $p_T$ and polar angle $\vartheta$
of the tracks are varied randomly. The fit function is given
by a double Gaussian. 
The momentum resolution is $\sigma(p)$/$p$=2.9\%.
} 
\label{ftrack}
\end{figure}

A helix track finder was developed for $\overline{\sf P}${\sf ANDA} \cite{thesis_muenchow}.
It was tested with Monte-Carlo simulated data for STT and MVD,
i.e.\ 30 plus $\leq$7 hits per track.
A field of $B_z$=2~T was used 
with field maps correctly treating overlap with the magnetic
dipole field in the $\overline{\sf P}${\sf ANDA} forward spectrometer.
The algorithm is based upon two steps.
In the first step, a conformal transformation is applied.
For every $x$,$y$ coordinate of hits in the STT or MVD,
new coordinates 
$x'$=($x$-$x_0$)/$r^2$ and 
$y'$=($y$-$y_0$)/$r^2$ 
with $r^2$=($x$-$x_0$)$^2$+($y$-$y_0$)$^2$ are calculated.
In a projection onto $x$$y$ plane, helix tracks are circles.
The conformal map transforms these circles
into straight lines, which can be indentified easier 
as tracks by a track finder.
In the second step, a Hough transform is applied.
For any combination of ($x$,$y$) coordinates a straight
line is formed, and the polar coordinates $r$ and $\theta$ 
are calculated. 
A normal vector with a 90$^0$ angle with respect 
to the line is constructed.
The parameter $r$ is the distance from ($x$=0,$y$=0) 
along the normal vector to the line,
and the parameter $\theta$ is the polar angle of the normal vector
in the $x$$y$ frame.
Then all the new coordinates are filled 
into a 2-dimensional ($r$,$\theta$) histogram,
and a peak finder is applied.
A peak in this histogram corresponds to a found track.
Fig~\ref{ftrack} (left) shows the Hough space for 
10 tracks of $p$=1~GeV/c.
The algorithm uses fix point arithmetics with 24 bit precision, 
in division and multiplication increased to 48 bit.
The size of the Hough space was adjusted to 512$\times$512.
The lookup table for the sinus function uses 128 values of 16 bit precision.
Fig~\ref{ftrack} (right) shows the momentum resolution for $p$=1~GeV/c tracks.
As a preliminary result \cite{thesis_muenchow}
the  efficiency of the online track finder 
is only $\simeq$20\% worse compared\footnote{The comparison between 
the online and the offline track finder algorithm was performed for
events containing 10 tracks with the same momentum, e.g.\ $p$=1~GeV/c,
but random variation of the $p_T$.} to the offline algorithm.
The $p_T$ resolution is only worse by a factor $\simeq$2.5.
For an online data reduction these values are acceptable.

\subsection{Event Selector Algorithm for HADES}

In order to test the speed of data moving on the CN,
an event selector algorithm was tested with HADES data
\cite{thesis_shuo}.
The algorithm was designed for 
{\it (a)} reading HADES binary events from DDR2 memory
{\it (a)} partially decoding the event, 
{\it (a)} issuing an accept or reject decision,
{\it (a)} discarding the event or writing it back to the DDR2 memory,
depending on which decision was issued.
For a DMA block size of 32 kB, 
for 100\% (25\%) accepted events the algorithm reached 
a throughput of $\simeq$80~MB/s ($\simeq$150~MB/s). 

\subsection{Additional Algorithms}

The matching of HADES tracks with the HADES time-of-flight and the 
HADES electromagnetic shower system requires track extrapolation 
through the $B$ field. 
As a preliminary result, for $^{40}$Ar+$^{40}$$[$KCl$]$ at 1.756 AGeV
a reduction of $\simeq$2 and an enhancement of $\simeq$1.8
was achieved at an efficiency of $\simeq$90\% \cite{talk_kopp}.
In addition, a track finder only based on hits of a silicon vertex detector
(i.e.\ 2 layers of a pixel detector and 4 layers of a strip detector)
was tested for the Belle II experiment \cite{thesis_muenchow}.

\subsection{Graphics Processing Units}

\label{c_gpu}

As a novel approach for fast data processing, a track fitter based upon
a conformal map transformation within the {\tt PandaRoot 2.0} 
framework \cite{pandaroot}
was tested on an NVidia Tesla C1060 Graphics Adapter \cite{mohammad}.
The card has 240 cores and a single precision floating point
performance of 933 GFLOPS.
For the calculations on the GPU (Graphics Processing Unit),
the NVidia CUDA framework \cite{cuda} was used and 
interfaced to {\tt PandaRoot}.
The syntax of CUDA is very similar to the ANSI C programming language.
The track finder for MVD and TPC was running in {\tt PandaRoot}
for tracks with generated $p$=1~GeV and 50-2000 tracks/event.
Then the hit data of the track candidates were transfered 
from the host PC to the GPU, where the track fitting was performed
in 32 parallel threads in the next step.
The fitted track data were transferred back to the PC.
The performance of the complete algorithm was compared 
between running with and without GPU (i.e.\ host PC alone).
A speed-up of a factor $\leq$68 \cite{mohammad} was achieved.
Thus, GPUs seem to be attractive solution for high level processing 
which require floating point operations and are not possible on an FPGA.

\acknowledgments

This work was supported by part by BMBF under contracts 06GI179 and 06GI180, 
GSI and DFG.

\end{document}